\documentclass{article}
\usepackage{amsmath}
\usepackage{graphicx,rotating}
\usepackage{psfrag}

\begin{document}
%-------------------------------------------------------------
% floats
%-------------------------------------------------------------

\renewcommand{\floatpagefraction}{1}
\renewcommand{\textfraction}{0}
\renewcommand{\topfraction}{1}
\renewcommand{\bottomfraction}{1}

%\documentstyle[german,a4,12pt,epsfig,citesort]{article}
%\documentstyle[a4,epsfig,12pt]{article} 
%\renewcommand{\baselinestretch}{1}
%\advance\hoffset by 0pt
%\advance\hsize by 0pt

\begin{center}
{\bf Self-organized Criticality on  Small World Networks}
\end{center}
\begin{center}
L. de Arcangelis$^1$  and H.J. Herrmann$^2$
%\footnote{On sabbatical leave from ICA1, University of Stuttgart.}
\end{center}
\begin{center}
Department of Information Engineering, Second University of Neaples$^1$\\
INFM Neaples UdR and CG SUN, Via Roma 29, I-81031 Aversa (CE), Italy\\ 
Institute for Computer Applications 1, University of Stuttgart$^2$\\
Pfaffenwaldring 27, D-70569 Stuttgart
\end{center}
\begin{abstract}
We study the BTW-height model of self-organized criticality on a
square lattice with some long range connections giving to the lattice
the character of small world network. We find that as function of the
fraction $p$ of long ranged bonds the power law of the avalanche size
and lifetime distribution changes following a crossover scaling
law with crossover exponents $2/3$ and $1$ for size and lifetime respectively.
\end{abstract}

Small world networks have recently been observed to exist in many
physical, biological and social cases \cite{Watts1998,Newman1999a,Newman1999b,Collins1998,Amaral2000}. A simple minded
example is the structure of the neo-cortex and the network of
acquaintances in certain  societies. Another ubiquitous phenomenon
(SOC) \cite{Bak1987,Bak-book,Manna1991}
in nature and society is self-organized criticality, i.e. the
appearance of avalanches of all sizes without characteristic scale over
a certain range. Easily one can imagine cases in which both of these
phenomena occur simultaneously, that means that we find an avalanche
type spreading on a sparsely long-range connected network. As an
example we propose the spreading of neural information inside the
cervical cortex which is due to the threshold behaviour given by the
firing rule which automatically induces avalanches of synapses. Another
example one can imagine are societies where each individual has a
certain threshold of endurance. Let us take the example of Peter
Grassberger for the classical SOC-model \cite{Bak1987} of office clerks moving
sheets of papers from desk to desk, as illustrated in Peter Bak's book
\cite{Bak-book} 
in Fig. 13 and admit that the clerks do not sit on a square lattice
but have a more realistic connectivity in their work relation
(typically small world network behaviour). Many other examples of this
kind can be thought of.

In this short paper we want to present a model for 
self-organized criticality on a graph having the properties of small
world networks. In fact, we consider the classical height model of Bak, Tang and
Wiesenfeld (BTW) \cite{Bak1987} on a square lattice which has been ``rewired''
for a certain fraction of bonds $p$ which are chosen of arbitrary range.  For
that purpose we take a square lattice of linear size $L$  
and all bonds present between nearest neighbours sites. 
Then we choose randomly two sites of the system
and place a bond between them (which can therefore be of any length
smaller than $\sqrt 2 L$). In order to keep the coordination of the
sites on average to be four, one of the smaller bonds going to a
neighbouring site of one of the end points of one long bond is removed.
This procedure repeats until a fraction $p$ of all bonds has been
replaced by long range connections. This type of graph is not the same
as the one used by most authors working on small world networks,
because the underlying short range lattice is not a linear chain but a
square lattice. Nevertheless the properties should qualitatively be
the same: large world behaviour for large $p$ and a small world 
behaviour for small $p$. The situation $p = 0$ corresponds to the simple 
square
lattice and $p = 1$ to random graph with average coordination 4 and
long range connections (Viana-Bray) on which one would expect
mean-field behaviour.

On each site of the lattice we place an integer value $h_{i}$
less than a threshold $h_c=6$. At each time step randomly one site $i$ 
is chosen and its value $h_i$ is increased by unity, i.e. a unit mass is 
added to it. 

When the value of the height $h_i$ having $q$ neighbours  reaches the 
threshold $h_c$, it topples,
i.e. it distributes mass equally to its neighbours:
\begin{eqnarray}
  h_{i} & \rightarrow & h_{i} - q  \\ 
  h_{j} & \rightarrow & h_{j} + 1 \notag
\end{eqnarray}

where $j$ goes over all $q$ neighbors of site $i$. We see that eq. (1)
preserves the mass (the sum of all heights).

We have studied the statistics of avalanches monitoring as well 
their size $s$ (number of sites that toppled at least once) 
as their lifetime $t$, that means the number of time step an avalanche
lives. The quantities $n(s)$ and $n(t)$ denote the avalanche size and 
life time distributions. For the case $p = 0$, that means the
classical BTW-model on the square lattice it is known that
asymptotically

\begin{eqnarray}
  n(s) &  \sim & s^{- \sigma} ,  \\  
  n(t) & \sim & t^{- \tau}   \notag
\end{eqnarray}
with $\sigma=1.0$ and $\tau=0.5$ in two dimensions.

It is our aim here to investigate what happens for different values of
$p$. To this purpose we have analysed square lattices of size $L=50, 100$
and $200$ and a range of $p$ values from $p=0$ to $p=0.5$. The lattice 
has spiral boundary conditions in one direction and two open boundaries at 
top and bottom, where mass in excess can flow out of the system. 
In each configuration we have injected randomly 2500 particles of unit mass, 
letting each time 
the avalanche proceed until no site had a value of $h>h_c$. The data were 
averaged over 200, 50 and 10 configurations for $L=50, 100$
and $200$ respectively.

\begin{figure}
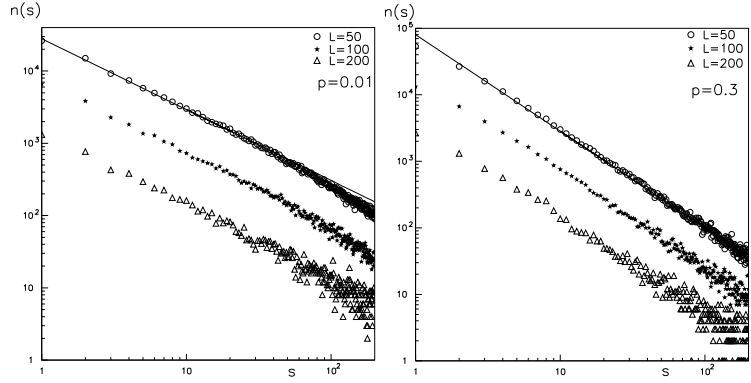

\begin{center}
\includegraphics[width=0.4\textwidth]{fig1a.epsi} 
\includegraphics[width=0.4\textwidth]{fig1b.epsi}
  \caption{Log-log plot of the distribution of avalanche size $n(s)$ as 
function of $s$ for different system size $L$ and for $p=0.01 (a)$ and 
$p=0.3 (b)$. The continuous lines indicate the slopes $0.98$ and $1.45$ for 
$(a)$ and $(b)$ respectively. 
The statistics is of 2500 particles injected in 200,50 and 10 configurations 
for $L=50,100$ and $200$ respectively.  }
  \label{fig:fig1}
\end{center}
\end{figure}

\begin{figure}
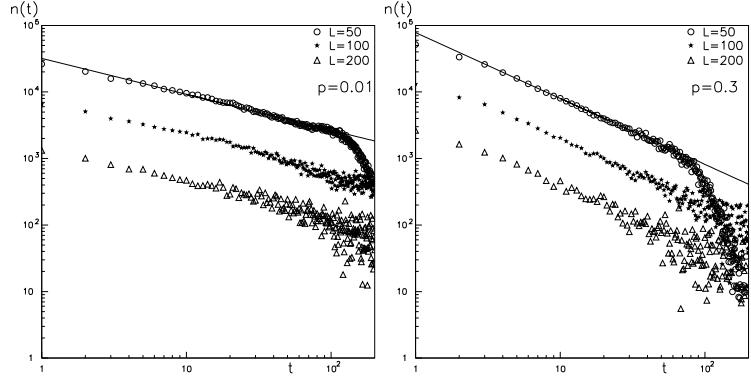

 \begin{center}
\includegraphics[width=0.4\textwidth]{fig2a.epsi}
\includegraphics[width=0.4\textwidth]{fig2b.epsi}  
\caption{ Log-log plot of the distribution of avalanche lifetime  $n(t)$ as
function of $t$ for different system size $L$ and for $p=0.01 (a)$ and
$p=0.3 (b)$. The continuous lines indicate the slopes $0.54$ and $0.99$ for
$(a)$ and $(b)$ respectively.
The statistics is the same as in Fig.~\ref{fig:fig1}.}
  \label{fig:fig2}
\end{center}
\end{figure}

In Fig. \ref{fig:fig1} we see the distribution of avalanche sizes for
different system sizes and two different values of $p, 0.1$ and $0.3$  
in a double
logarithmic plot. We see that the data follow a straight line for
nearly two decades indicating that we still find SOC-behaviour. The 
slope of the straight line gives us the exponent
$\sigma$. In Fig.~\ref{fig:fig2} we see the corresponding figures for the
distribution of life times. Again the data show power law behaviour
and the slope gives the exponent $\tau$. We observe that the exponents
$\sigma$ and $\tau$ depend on $p$ but they do not appear to depend on $L$. 

\begin{figure}
\begin{center}

\includegraphics[width=0.4\textwidth]{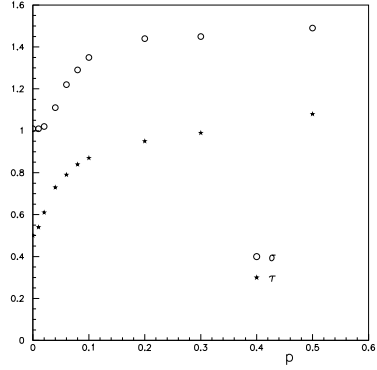}
  \caption{ Critical exponents $\sigma$ and $\tau$ for the distribution of 
avalanche size and lifetime as function of $p$. The exponent are calculated for the system size $L=50$. }
  \label{fig:fig3}
\end{center}
\end{figure}

\begin{figure}
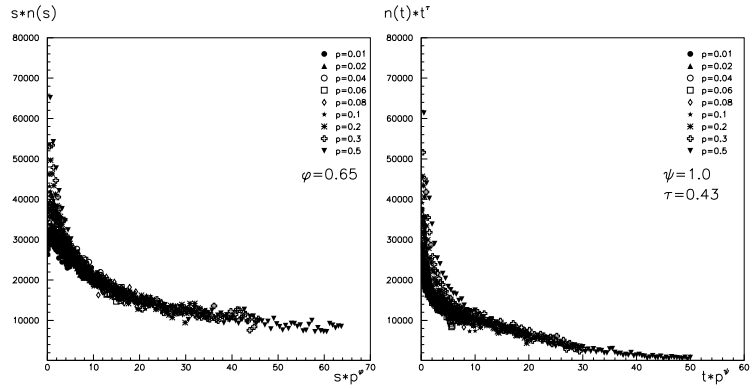

\begin{center}  
\includegraphics[width=0.4\textwidth]{fig4a.epsi}
\includegraphics[width=0.4\textwidth]{fig4b.epsi}
\caption{Data collapse of the distributions of avalanche size $n(s)$ 
and lifetime $n(t)$ for different values of $p$ and $L=50$. 
$(a)$ The plot of the quantity $n(s)s^{\sigma}$ for ${\sigma}=1.0$ as function 
of $sp^{\phi}$ gives a value for the crossover exponent $\phi=0.65\pm 0.1$. 
$(b)$ The analogous plot of the quantity $n(t)t^{\tau}$ for ${\tau}=0.5$
as function
of $tp^{\psi}$ gives a value for the crossover exponent $\psi=1.0\pm 0.1$. }
  \label{fig:fig4}
\end{center}
\end{figure}

In Fig.~\ref{fig:fig3} we see the dependence of $\sigma$ and $\tau$ on
$p$ as
obtained from our simulations. We see that for $p \to 0$ we obtain the
classical result on the square lattice of BTW  $\sigma = 1$, $\tau
= 1/2$ \cite{Bak1987}. For $p$ close to unity the values of the exponents 
convert to
the meanfield values $\sigma = 3/2$ and $\tau = 1$ \cite{Tang1988,Janowsky1993,Flyvbjerg1993}. This is of course
not surprising and the question is if the continuous change of the
exponent $\sigma$ and $\tau$ is an intrinsic continuous line of
critical points or, if we have here a crossover phenomenon as it appears
for instance in magnetic models that interpolate between 2 and 3
dimensions. For that purpose we tried for both the size and the
lifetime avalanche data in Fig.~\ref{fig:fig4}, a data collapse of all 
the distributions for
different values of $p$ following the classical crossover scaling
\begin{eqnarray}
  n_{p}(s) &  = & s^{- \sigma} {\cal{F}} (sp^{\phi}) \\ 
\nonumber \\
 n _{p}(t) & = & t^{- \tau} {\cal{G}} (tp^{\psi})  \notag
\end{eqnarray}
where $\cal{F}$ and $\cal{G}$ are scaling functions and $\phi$ and
$\psi$ are universal crossover exponents. We see from Fig.~\ref{fig:fig4} that a collapse of the data works reasonably well
yielding  crossover exponents of $\phi = 0.65 \pm 0.1$ for the avalanche size 
distribution and $\psi = 1.0 \pm 0.1$ for the lifetime distribution.

By studying the BTW-model on a small world square network we observed
a crossover to meanfield behaviour following a crossover scaling law
of eq. (3). It would be interesting to see if the same occurs for the simpler
Manna-model and we have heard calculations are already under way \cite{St}.

This work has been partially supported by the European TMR Network-Fractals
under contract No.FMRXCT980183 and by MURST-PRIN-2000.

\newpage

\begin{thebibliography}{10}

\bibitem{Watts1998}
Watts, D.J., Strogatz, S.H.
\newblock{Collective dynamics of `small-world' networks.}
\newblock Nature 393, 440--442 (1998).

\bibitem{Newman1999a} 
Newman, M.E.J., Moore, C., Watts, D.J.
\newblock {Mean-Field Solution of the Small-World Network Model.}
\newblock Physical Review Letters 84, 3201--3204 (1999).

\bibitem{Newman1999b} 
Newman, M.E.J., Watts, D.J.
\newblock {Scaling and percolation in the small-world network model.}
\newblock Physical Review E {\bf 60}, No. 6, 7332--7342 (1999).

\bibitem{Collins1998} 
Collins, J.J., Chow, C.C. 
\newblock {It's a small world.}
\newblock Nature 393, 409 (1998)

\bibitem{Amaral2000} 
Amaral, L.A.N., Scala, A., Barth\'el\'emy, M., Stanley, H.E.
\newblock {Proc.Nat.Ac.Sci USA 97, 11149-11152 (2000).}

\bibitem{Bak1987} 
Bak, P., Tang, C. and Wiesenfeld K.
 \newblock {\em Self-Organized Criticality. An Explanation of $1/f$ Noise.}
\newblock Physical Review Letters 59, 381 (1987).
\newblock {\em Self-Organized Criticality.}
\newblock Physical Review A {\bf 38}, 364 (1988).

\bibitem{Bak-book} 
Bak, P.
\newblock {How nature works. The science of self-organized criticality.}
\newblock Springer, New York (1996).

\bibitem{Manna1991}
Manna, S.S.
\newblock J. Phys. A {\bf 24}, L363 (1991).

\bibitem{Tang1988}
Tang, C., Bak, P.
\newblock J.Stat.Phys. {\bf 51}, 797 (1988).

\bibitem{Janowsky1993}
Janowsky, S.A., Laberge, C.A.
\newblock J. Phys. A {\bf 26}, L973 (1993).

\bibitem{Flyvbjerg1993}
Flyvbjerg, H., Sneppen, K., Bak, P.
\newblock Phys. Rev. Lett. {\bf 71}, 4087 (1993).

\bibitem{St}
Stanley, H.E. et al.\\
\newblock private communication.
\end{thebibliography}
\end{document}